# Realization of Non-diffracting and Self-healing Optical Skyrmions

A. Srinivasa Rao[1,2,3*]

[1]Graduate School of Engineering, Chiba University, Chiba, 263-8522, Japan
[2]Molecular Chirality Research Centre, Chiba University, Chiba, 263-8522, Japan
[3]Institute for Advanced Academic Research, Chiba University, Chiba, 263-8522, Japan
E-mail: [*]asvrao@chiba-u.jp, sri.jsp7@gmail.com

**Abstract**
Optical skyrmions formed in terms of polarization are topological quasi-particles and have garnered much interest in the optical community owing to their unique inhomogeneous polarization structure and simplicity in their experimental realization. These structures belong to the Poincaré beams satisfying the stable topology. We theoretically investigated the non-diffracting and self-healing Poincaré beams based on the superposition of two orthogonal Bessel modes by mode matching technique. These Poincaré beams are topologically protected, and we suggest them as optical skyrmions.  These optical skyrmions are quasi-skyrmions and their range of propagation depends on the range of superposed Bessel modes. The polarization structure of these optical skyrmions has no change upon the propagation. A necessary experimental configuration is suggested to realize variable order skyrmions in Bessel modes experimentally. This work can provide a new direction for the generation of skyrmions with completely new textures and features with respect to existing skyrmions originating from Laguerre-Gaussian modes.

Keywords: optical skyrmions, skyrmion number, Poincaré beams, Bessel beams, non-diffraction, self-healing, vector beams, Poincaré Bessel beam

## 1. Introduction

A skyrmion is a topologically protected quasiparticle generally formed by the collection of *n* number of entities from a single group [1]. The skyrmionic textures can be quantitatively understood by the associated skyrmion number $N_{sk}$. The topological textures of skyrmions have been observed and well-studied in multiple science fields, predominantly in particle physics [2], condensed matter physics [3], magnetism [4], string theory [5], spintronics [6], Bose-condensates [7], plasmonics [8], quantum hall systems [9], and photonics [10].

In recent years, skyrmions were also manifested in laser modes by considering polarization as an entity and these laser modes are often termed optical skyrmions [11,12]. The optical skyrmions belong to the family of full Poincaré beams and can be distinguished from others of the family by their topological protection and skyrmion number [13,14]. The skyrmion number can be easily understood and estimated by a stereographic projection of the three-dimensional (3D) polarization distribution in the Poincaré sphere onto a 2D plane.  The 2D plane is the transverse plane of the Poincaré beam. Therefore, the skyrmion number associated with any optical skyrmion can be obtained in terms of stokes vector $\vec{S} = [S_1, S_2, S_3]$ confined in a spherical surface area $\sigma$ [15] as

$$N_{sk} = \frac{1}{4\pi} \iint_{\sigma} \vec{S} \cdot \left( \frac{\partial \vec{S}}{\partial x} \times \frac{\partial \vec{S}}{\partial y} \right) dx dy. \tag{1}$$

It is worth noticing that the skyrmion number provides the number of times the tip of the vector covers the entire sphere. These skyrmion modes have integer skyrmion numbers and they must be conserved throughout their propagation. The skyrmion beams can be experimentally realized easily by the superposition of two orthogonal spatial modes that have orthogonality in their circular polarization.

$$|F(r,z)\rangle = \frac{1}{\sqrt{2}}[f_R(r,\phi,z)|R\rangle + exp(i\theta_0)f_L(r,\phi,z)|L\rangle]. \tag{2}$$

Here, $f_R(r,\phi,z)$ and $f_L(r,\phi,z)$ are the wave functions of two orthogonal spatial eigenmodes. The ket vectors: $|R\rangle$ and $|L\rangle$ represent respective right and left circular polarization states. The global phase difference between the two modes is denoted by $\theta_0$. We can easily fabricate all types of skyrmionic textures, *viz*. Néel-type and Bloch-type skyrmions and anti-type skyrmions in laser modes by simply controlling the phase difference $\theta_0$ and angle of a quarter-wave plate in any kind of generation scheme. The Néel-type and Bloch-type skyrmions have the respective hedgehog and vortex textures while the anti-skyrmions have the saddle texture.

Further, we can understand from the mathematical representation of optical skyrmion given in Eq. 2, that to realize the skyrmion texture in laser beams, we just need two cylindrically symmetric eigenmodes for crafting optical skyrmion textures. Moreover, this necessity was successfully fulfilled by Laguerre Gaussian (LG) modes owing to their three properties. The first one is their cylindrical symmetry which can easily allow us to do the stereographic projection between the 2D plane and the Poincaré sphere. The second property is the order-dependent change in their radial intensity which provides the uniform transformation between the two orthogonal handedness of polarizations through weight factors. The third property is their order-dependent helical wavefront which contributes to a desired azimuthal phase difference between the two orthogonal circular polarizations to produce the orientation in the elliptical polarization. As given in Eq. 3, we can replace $f_R(r,\phi,z)$ and $f_L(r,\phi,z)$ wave functions with amplitudes $E_\ell$ of two orthogonal LG modes $LG_0^\ell$ in Eq. 2 to generate LG beam based skyrmion texture.

$$|\Psi(r,z)\rangle = \frac{1}{\sqrt{2}}\left[E_{\ell_R}(r,\phi,z)|R\rangle + exp(i\theta_0)E_{\ell_L}(r,\phi,z)|L\rangle\right], \qquad (3)$$

with satisfying the condition $\ell_L \neq \ell_R$. The Gouy phase difference created by virtue of two orthogonal LG modes changes the state of the polarization in the skyrmion as it propagates, however, the skyrmion number remains conserved. The skyrmion textures can be easily constructed in LG modes by several techniques. For instance, superposition of LG modes, generated from SLM, in sagnac interferometer can produce variable order skyrmions [16], double pass of Gaussian mode to a single SLM loaded with desired phase hologram can also deliver variable order skyrmions [17], the skyrmionic textures made by the spin-orbit coupling of tightly-focused non-paraxial optics [18], we can create multi-Skyrmions in single beam by the superposition of two non-zero OAM modes of LG family [19], arbitrary skyrmions generation directly from optical micro-ring cavity [20].

In recent times, multiple attempts have been made as a forward step in deep investigation and understanding of the properties of LG modes based optical skyrmions towards their applications. For example, investigation of mode mismatching in paraxial skyrmions under focusing conditions [21], deep understanding of skyrmion number of paraxial modes [12,19], disorder-induced topological state transitions in optical skyrmions [22], skyrmion-like structures formed by Poynting vectors are investigated in the focal region of a pair of counter-propagating cylindrical vector vortex beams in free space [23], a study of spin vortices and skyrmions coherently imprinted into an exciton-polariton condensate on a planar semiconductor microcavity [24], spin-orbit interaction of skyrmions at optical interface [25].

Here, we first show that the generation of Poincaré modes in Bessel profiles based on the pump mode matching technique. The analysis was further utilized to demonstrate a single type of skyrmion without mode transformation between Néel-type and Bloch-type states while it is propagating based on mode matching technique generally governed by pump beam transverse intensity. On top of that, we suggest a simple and cost-effective experimental configuration that can produce our proposed skyrmion textures with high output power even in ultra-short laser pulses. Based on our method we can generate variable-order optical skyrmions in the Bessel host. The non-diffraction and self-healing skyrmions can have potential applications in light-matter interactions.

## 2. Theory

In addition to LG modes, we have another set of cylindrical symmetric eigenmodes with helical wavefronts which are called Bessel modes [26]. The requirements for the skyrmion texture generation discussed in the introduction can be fulfilled by the Bessel beams. Therefore, an intriguing curiosity arises about whether it is possible to create skyrmionic texture in Bessel profiles. For this purpose, first, we have to produce Poincaré beams in Bessel profiles and then investigate whether these structures are topologically stable to form skyrmions. As part of this, first of all, from Eq. 2, we can obtain the wave function for the skyrmion in terms of Bessel modes as

$$|\varphi(r,z)\rangle = \frac{1}{\sqrt{2}}\left[B_{\ell_R}(r,\phi,z)|R\rangle + exp(i\theta_0)B_{\ell_L}(r,\phi,z)|L\rangle\right]. \qquad (4)$$

Here, $B_\ell$ is the $\ell^{\text{th}}$ order ideal Bessel beam and is given by

$$B_\ell(r,\phi,z) = exp(ik_z z - i\omega t)J_\ell(k_r r)exp(i\ell\phi). \qquad (5)$$

As shown in Eq. 5, it is very important to note that the ideal Bessel beam has a constant transverse profile throughout its propagation, i.e., independent of longitudinal position *z*. Hence, they can provide skyrmions which can have invariant topological charge throughout their propagation. However, this is not true when it comes to the experiments because it is not possible to experimentally realize infinite width and propagation range Bessel beams which are necessary conditions for an ideal Bessel beam. Indeed, we can experimentally generate Bessel beams with finite width and propagation range, which we often call quasi-Bessel beams, based on wavefront division interference through several methods [27-29]. Consequently, we can generate only quasi-skyrmions in real Bessel beams. On top of that, the mode matching between the two interfering Bessel beams must be constant throughout the propagation. Besides, the longitudinal intensity distribution of Bessel modes strictly depends on their order. Hence, it is a challenge to generate two Bessel beams with different order but their longitudinal profiles are identical for realization of Bessel skyrmions. This scenario can be straightforwardly visualized in Bessel beams which are generated by pumping the so-called Gaussian vortex mode $LG_0^\ell$ (LG mode with zero radial index) to the axicon. The amplitude of the Gaussian vortex mode is given by [30]

$$E_\ell(r) = \sqrt{\frac{2^{\ell+1}P}{\ell!\,\pi w^2}} \left(\frac{r}{w}\right)^\ell exp\left(\frac{-r^2}{w^2}\right) exp(i\ell\phi), \tag{6a}$$

here, $P$ is the total optical power carried by the Gaussian vortex beam with Gaussian spot size $w$ and $\ell$ is the topological charge of the LG mode. The intensity distribution of the Gaussian vortex mode is given by

$$E_\ell^2(r) = \frac{2^{\ell+1}P}{\ell!\,\pi w^2} \left(\frac{r}{w}\right)^{2\ell} exp\left(\frac{-2r^2}{w^2}\right). \tag{6b}$$

The intensity distribution of the Bessel beam formed by pumping the axicon with Gaussian vortex mode can be mathematically expressed as [31]

$$B_\ell^2(z,r) \propto \left(\frac{z}{z_{max}^G}\right)^{2\ell+1} exp\left(\frac{-2z^2}{{z_{max}^G}^2}\right) J_\ell^2(k_r r). \tag{7}$$

Here, $k_r = sin[(n-1)\alpha]$ is the radial component of the propagation vector and $z_{max}^G = kw/k_r$ is the range of the Bessel beam formed in the presence of a Gaussian beam as a pump source with $\alpha$ and $n$ as the base angle and refractive index of the axicon. From Eqs. 6 and 7, we can easily infer that the order of the Gaussian vortex mode converted into the order of the corresponding Bessel beam with invariable transformation. As well, the transverse intensity profile of pump mode is projected onto the longitudinal axis to produce a Bessel beam. Thus, the longitudinal intensity distribution of the Bessel beam strictly depends on the transverse intensity distribution. In Fig. 1, we can see that the range and position of the Bessel beam are changing with its order in the way the change occurred in the transverse intensity profile of the Gaussian vortex beam. The inner and outer radii of the Gaussian vortex mode, represented with $r_1$ and $r_2$ respectively, constrain the longitudinal intensity distribution of the Bessel beam. Indeed, the onset and offset positions of the Bessel beam of order $\ell$ can be written in terms of its pump Gaussian vortex radii as $z^\ell_{min} = r_1/(n-1)\alpha$, and $z^\ell_{max} = r_2/(n-1)\alpha$ [32]. Also, the range and peak position of the Bessel beam along its propagation direction is $z^\ell_{range} = (r_2 - r_1)/(n-1)\alpha$ and $z^\ell(I_{max}) = z^G_{max}(2\ell+1)^{1/2}/2$ respectively.

Past few years, three groups have successfully demonstrated the Poincaré modes in Bessel profiles. The first one is V. Shvedov et al. [33] created Poincaré modes by perfectly matching the 0[th] and 1[st] order Bessel modes based on conical diffraction and Bessel beam formation with a high-optical quality biaxial crystal [34]. Other two groups have used the superposition of two Bessel beams with different longitudinal propagation vectors to generate Poincaré modes in Bessel beams [35,36]. It is also worth noticing that this technique can produce an optical bottle beam with non-diffraction and self-healing properties under the same state polarization attributed to differences in the longitudinal propagation vectors [37,38].

Recently, K. Singh et al. [39] have successfully demonstrated quasi-skyrmions by using the superposition of two Bessel modes with different propagation vectors used in the references [37, 38]. The Gouy phase acquired due to the difference in the propagation vector allows them to produce skyrmions whose propagation characteristics are similar to the skyrmions generated in LG modes. As a result of the Gouy phase, the transformation between Néel-type and Bloch-type skyrmions takes place with propagation. For a quantitative understanding of these skyrmions, they derived a mathematical expression for the Skyrmion number, and it is given by

$$N_{sk} = q\Delta\ell \left[\frac{1}{1+\alpha(0)^2} - \frac{1}{1+\alpha(r_0)^2}\right] \tag{8}$$

with $\alpha(r) = \left|B_{\ell_L}(rk_r)/B_{\ell_R}(rk_r)\right|^q$, $\Delta\ell = \ell_R - \ell_L$, and $q = (|\ell_L| - |\ell_R|)/||\ell_L| - |\ell_R||$. The Skyrmion number extracted by using Eq. 8 is in the localized circular region bounded by the condition $r \leq r_0$.

The alternative way to generate skyrmions in the Bessel profile is by perfectly longitudinally mode matching. This can be achieved by properly modulating the initial beam used in Bessel mode conversion. Here the two superposed modes have the same propagation vector so there will not be any transformation between Néel-type and Bloch-type Skyrmion textures along the propagation.

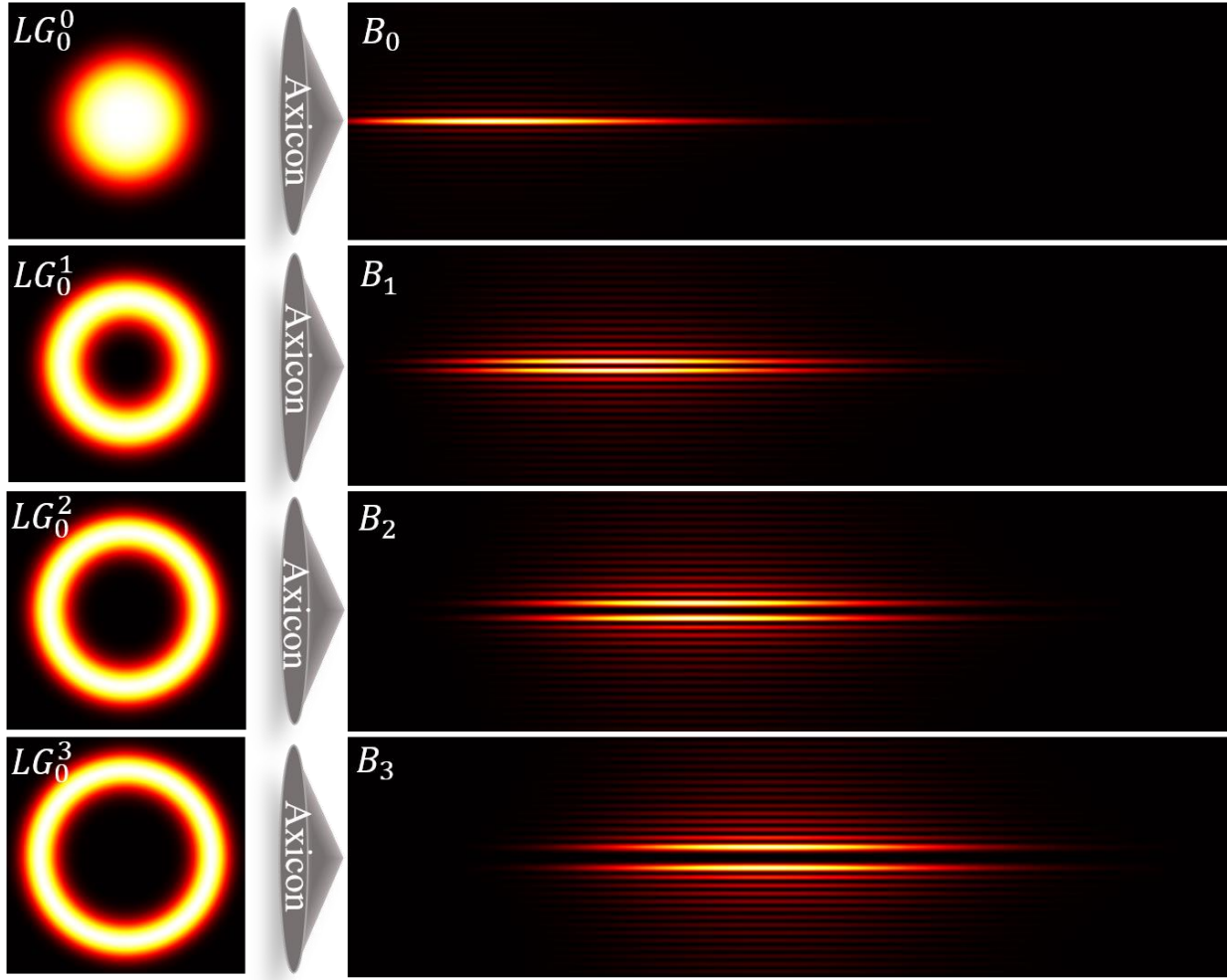

Fig. 1. Mode conversion of Gaussian vortex into Bessel in the presence of axicon.

### 3. Optical skyrmions generation by the superposition of $0^{th}$ order and higher order Bessel beams

As shown in Fig. 1, the range and position of the higher order Bessel beam change as a result of a change in the doughnut size of the Gaussian vortex mode. We can match the higher order Bessel modes with the $0^{th}$ order Bessel mode by pumping axicon with doughnut shape intensity distribution without any helical wavefront. We can fulfill this requirement with a Hollow Gaussian (HoG) beam. It has a doughnut-shaped intensity distribution with no helical wavefront. This beam was introduced by Y. Cai in 2003 and the mathematical Eq. is [40]

$$E_m(r) = \sqrt{\frac{2^{2m+1}P}{(2m)!\,\pi w^2}} \left(\frac{r^2}{w^2}\right)^m exp\left(\frac{-r^2}{w^2}\right), \qquad (9)$$

here, $m$ is the order of the HoG beam. HoG mode can be easily produced in the experimental laboratory by suitable computer-generated hologram projected onto a spatial light modulator [41,42], by utilizing the thermal lens effect [43], and by using two spiral phase plates with the same order arranged in opposite directions with each other [44,45]. The HoG-pumped axicon produces the $0^{th}$ order Bessel mode whose range and peak position depend on the order of the corresponding HoG mode [46]. The intensity distribution of $0^{th}$ order Bessel beam generated by pumping the axion with HoG mode provided by Eq. 9 is given by

$$B_0^2(z,r) \propto \left(\frac{z}{z_{\max}^G}\right)^{4m+1} exp\left(\frac{-2z^2}{z_{\max}^{G\,2}}\right) J_0^2(k_r r). \tag{10}$$

From Eqs. 7 and 9, we can infer that the order of HoG is double the order of Gaussian vortex and we can mode match only even order vortex modes with the HoG for generating invariable Poincaré modes in Bessel profiles. However, C. Wei et al. introduced a method to generate order HoG modes from Gaussian vortex modes with the condition $m = \ell/2$ [47]. Hence we could generate HoG modes that have one-to-one mode matching with all the Gaussian vortex modes. Therefore, we can have a way to tailor order tunable skyrmions in Bessel Poincaré beams via stokes vector fields by pumping suitable order HoG and Gaussian vortex modes to a single axicon.

As depicted in Fig. 2, by using a Mach-Zander Interferometer (MZI) and three mode converters we can generate all orders of optical skyrmions in Bessel profiles by setting up one of $\ell_R$ and $\ell_L$ to zero and the other one as variable. A linearly polarized and collimated Gaussian beam from the laser source can deliver two orthogonally linear polarized Gaussian modes with equal intensity with a half-wave plate ($\lambda/2$) and first Polarizing Beam Splitter ($PBS_1$). Further, we can convert these Gaussian modes into HoG mode in one arm and LG mode in the other arm of MZI. These two linearly orthogonal polarized modes can be superposed by the $PBS_2$. The polarization basis of the superposed state can be transformed from linear to circular in the presence of a quarter-wave plate ($\lambda/4$), whose fast axis is at $\pm 45°$ with respect to horizontal. By passing the final superposed state through an axicon we can generate variable order skyrmions and anti-skyrmions in terms of order of HoG and LG modes. It is worth noticing that the two Bessel beams generated in the superposition state have identical propagation vector spectrum. Hence, there will not be any propagation vector phase mismatch in the skyrmions. The polarization texture can be transformed between Néel-type and Bloch-type skyrmions by providing phase delay $\theta_0$ through the delay stage formed by the $DM_i$ mirrors. Thus, we can successfully generate all the skyrmions presented in Table 1. The sign and magnitude of the skyrmion number are given by the product $q\Delta\ell$. In the suggested experimental configuration under spiral phase plates and an axicon as mode converters, we can generate high power optical skyrmions in laser beams with maximum conversion efficiency. Further, it can be directly used for the generation of skyrmions in high-peak power laser pulses.

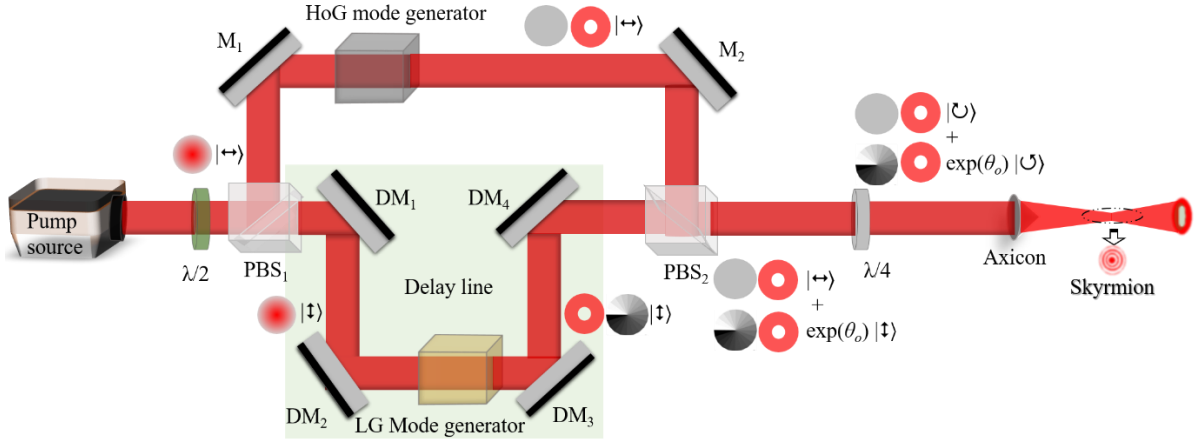

Fig. 2. Schematic experimental configuration can be used for the generation of variable order optical skyrmions in Bessel profiles. Here, $PBS_i$ is the polarizing beam splitter, $M_i$ is the Mirror, $DM_i$ is the delay line Mirror, $\lambda/2$ is the half-wave plate, and $\lambda/4$ is the quarter-wave plate. Double and circular arrows represent the polarization direction.

Table. 1. Types of skyrmions and their properties in terms of individual orthogonally polarized scalar Bessel modes involved in the superposition. RCP and LCP are referring to right circular polarization and left circular polarization.

| $\ell_R$ | $\ell_L$ | Polarization distribution | $N_{sk}$ | Skyrmion type |
|---|---|---|---|---|
| 0 | $+\ell$ | Leman, monstar, and flower structures with RCP at stokes vortex $S_{12}$ | $-\ell$ | Néel-type and Bloch-type |
| $-\ell$ | 0 | Leman, monstar, and flower structures with LCP at stokes vortex $S_{12}$ | $+\ell$ | |
| 0 | $-\ell$ | Star, and spider web structures with RCP at stokes vortex, $S_{12}$ | $+\ell$ | anti-type |
| $+\ell$ | 0 | Star, and spider web structures with LCP at stokes vortex, $S_{12}$ | $-\ell$ | |

The longitudinal cross-section of Bessel skyrmions of the first three orders which can be experimentally realized by the experimental configuration given in Fig. 2 are depicted in Fig. 3. The size and shape of $HoG_m$ and $LG_0^\ell$ modes used in the superposition are identical. However, the HoG$_m$ mode has a constant transverse phase while the LG$_0^\ell$ mode has a helical wavefront in its transverse profile. Here, the 0$^{th}$ order Bessel beam is created by the HoG$_m$, and $\ell^{th}$ order Bessel beam is delivered by the LG$_0^\ell$ mode.

The transverse intensity and polarization distributions of the first three skyrmions and anti-skyrmions are presented in Fig. 4. Here, the red and blue polarization ellipses indicate the right-handed and left-handed polarization. The intensity mismatch between the two superposed Bessel beams taking place in the radial direction creates periodic polarization change between right circular and left circular polarizations. More interestingly, the transverse profile exhibits polarization singularities of C-points and L-lines owing to the difference in the azimuthal phase and radial intensity of two Bessel beams. Indeed, the central parts of the two superposed Bessel beams produce skyrmion while their outer rings provide a combination of two equal orthogonal elliptical polarizations separated with *L*-lines [48]. The protected skyrmion structures by outer Bessel rings are white-circled. The effective skyrmion number of the outer rings calculated with Eq. 8 is zero. These skyrmions always have central bright intensity irrespective of their order and can be called bright skyrmions. The self-healing nature of skyrmions bestowed from their parent Bessel beams. The same transverse profile of pump modes to the axicon produces a constant weight factor ($\gamma_{\ell,\ell'}$) which is independent of spatial coordinates. The weight factor in terms of individual Bessel beams can be expressed as $\gamma_{\ell,\ell'} = B_\ell(r,z)|_{peak} / B_{\ell'}(r,z)|_{peak}$ and $B_\ell(r,z)|_{peak}$ is peak intensity of $\ell^{th}$ order Bessel beam along its propagation. The constant weight factor produces a fixed radial position for circular *L*-lines. Therefore, the size of the skyrmions is constant along its propagation and they become non-diffraction.

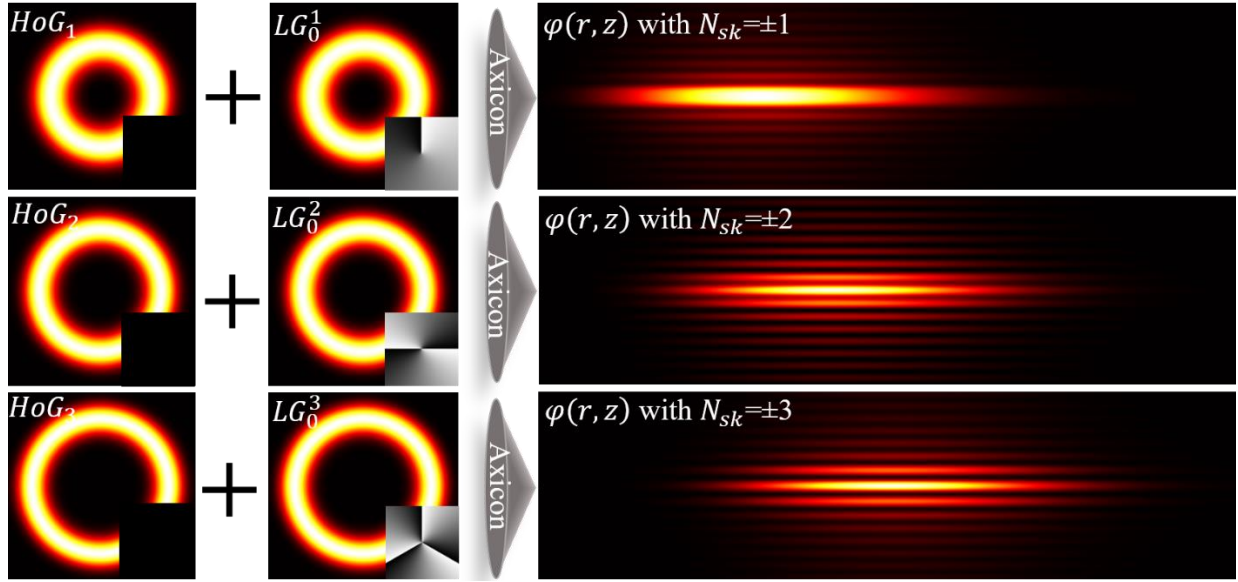

Fig. 3. Optical skyrmions are generated in Bessel profiles by pumping the axicon with a superposed mode formed by the hollow Gaussian mode and Gaussian vortex mode. The phase profile of hollow Gaussian and Gaussian vortex modes is provided as an inset in the corresponding mode.

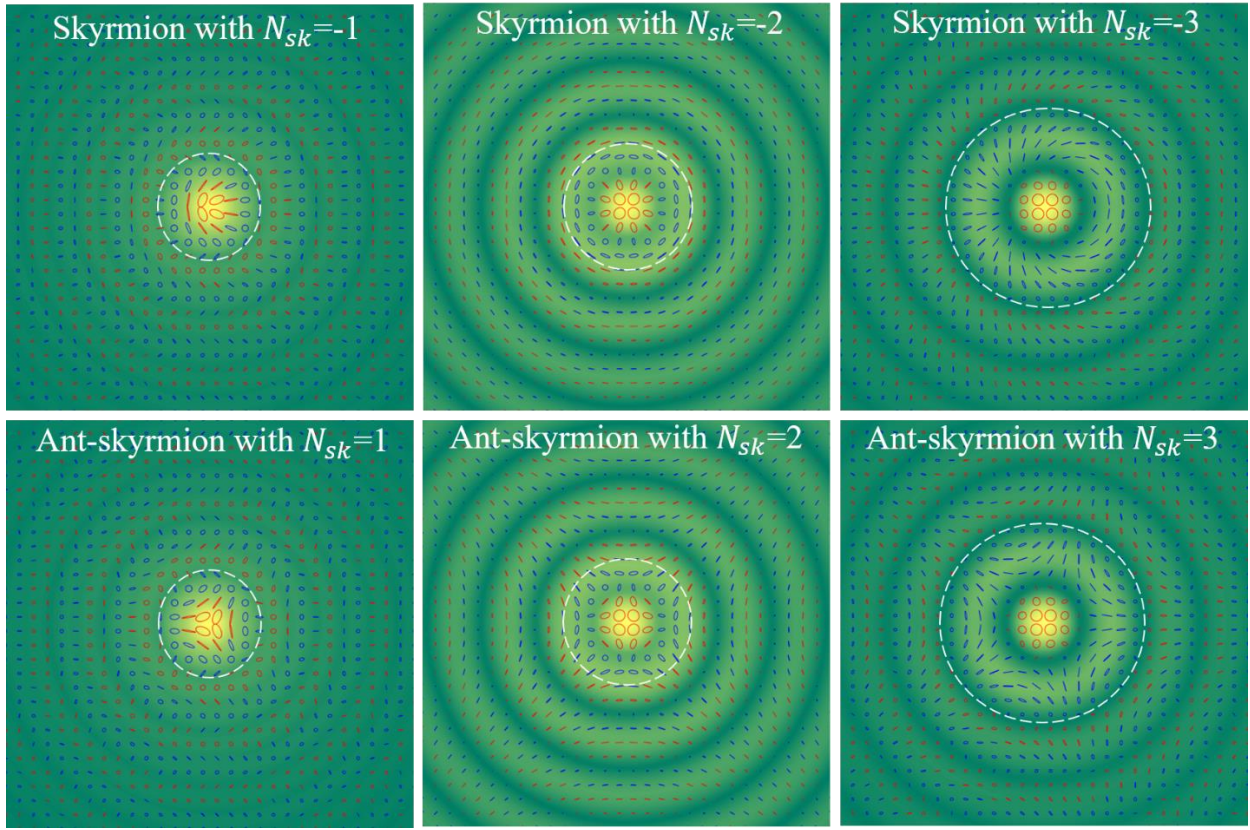

Fig. 4. The transverse profile of optical skyrmion textures created in Bessel beams by superposition of $0^{th}$ and $\ell^{th}$ order Bessel beams: Néel-type skyrmions are given in the first row and anti-skyrmions provided in the second row.

## 4. Optical skyrmions generation by superposition of two higher order Bessel beams

Similar to the preceding section, we can generate optical skyrmions in the Bessel profile by the superposition of two higher order Bessel beams. As discussed in Fig. 1, the position of Bessel beams along the propagation depends on their order. Hence, it is not possible to generate these skyrmions by directly using the experimental configuration shown in Fig. 2. However, by using two LG mode converters and two identical axicons in the two arms of MZI, we can generate two Bessel beams with identical propagation vectors independently. Furthermore, we can match the peak position of the two higher order Bessel beams by positioning one of the axicons on a longitudinal translation stage. The skyrmion textures of different orders at their peak intensity position are presented in Fig. 5. The polarization distribution of skyrmions is separated from the outer polarization distribution with the white circle. These skyrmions have a central dark core at their center owing to the higher order Bessel beams participation in the superposition and can be called dark skyrmions.

The propagation characteristics of the dark skyrmions are different from those of bright skyrmions in the present context. The propagation characteristics can be understood through the weight factor $\gamma_{\ell,\ell'}$. We presented the spatial dependence of the weight factor along the propagation in Fig. 6. Here, we normalized the intensity distributions of Bessel beams with their peak intensity, and we achieved the weight factor of ideal skyrmions in the Bessel profile ($\gamma_{ideal}$) is one. The position $z = 0$ corresponds to the peak intensity position of the Bessel beam. In the superposition of higher order Bessel beams, even though we perfectly match their peak position there will be a certain longitudinal mode mismatch occurs and this is anticipated since the longitudinal intensity distribution of Bessel beams depends on their order. The mode mismatch increases while we are moving on either side of the peak intensity position. Hence, we could have a skyrmion structure only in certain regions around the peak position. It is worth noting that the mode mismatch decreases with decreasing $\Delta\ell = |\ell_R - \ell_L|$ and this mismatch is less for higher order Bessel modes. Hence, it is best practice to use the maximum possible order of Bessel beams to increase the skyrmion propagation range.

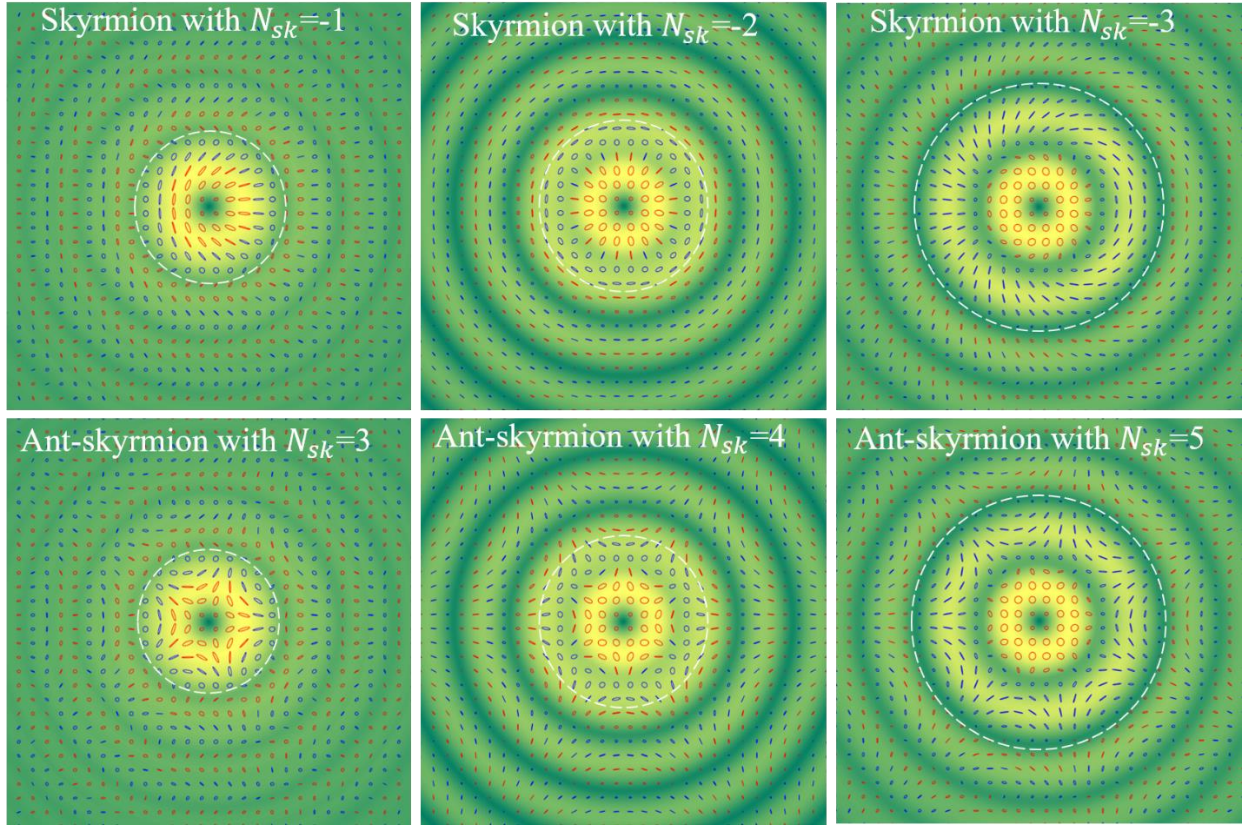

Fig. 5. The transverse profile of optical skyrmion textures created in Bessel beams by superposition of two higher order Bessel beams: Néel-type skyrmions are given in the first row and anti-skyrmions provided in the second row.

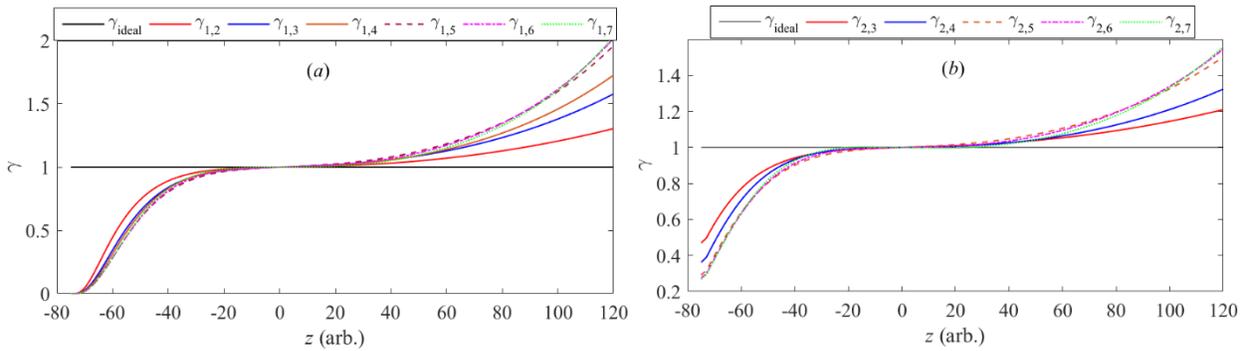

Fig. 6. Weight factor ratio of higher order Bessel beams involved in the skyrmions generation (*a*) with respect to first order Bessel beam and (*b*) with reference to second order Bessel beam.

To further understand the effect of longitudinal mode mismatch on the transverse profile of skyrmions, we have investigated the transverse polarization distribution as a function of longitudinal position for the superposition of 1st order and 4th order Bessel beams. The results are illustrated in Fig. 7. The polarization distribution in the transverse plane changes with the longitudinal position. In the $z < 0$ region, left-handed polarization is dominating, and in the $z < 0$ region, right-handed polarization is enhanced. As a result, the outer rings have a nonzero skyrmion number.

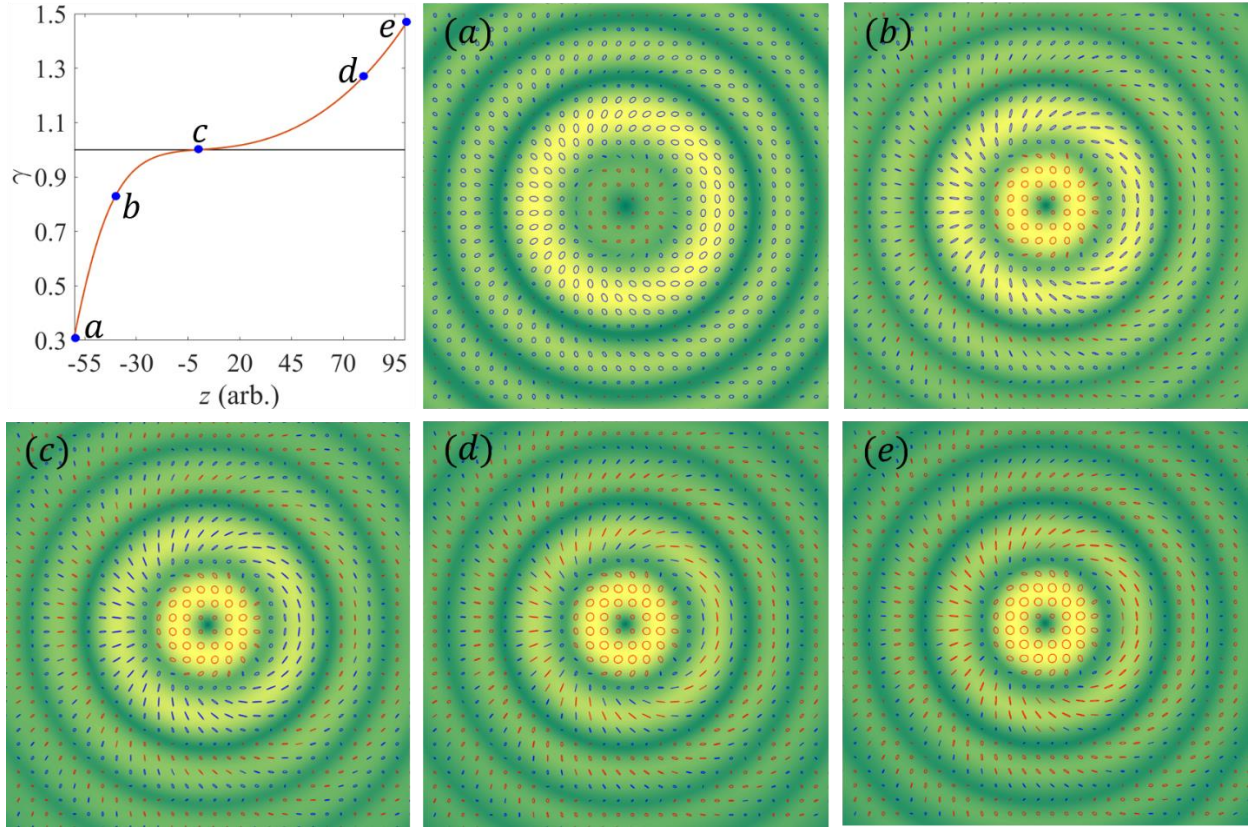

Fig. 7. Effect of weight factor ratio on the propagation of third order skyrmion which is formed by the superposition of 1st order and 4th order Bessel beams. Transverse profile of third order skyrmion at *a*, *b*, *c*, *d*, and *e* positions are given in the order they occurred in the propagation.

## 5. Discussion

Even though Bessel rings have no direct contribution to the skyrmion texture, they play a pivotal role in the reconstruction and non-diffraction of skyrmions. Indeed, the non-diffraction and self-healing ranges of skyrmions are directly proportional to the number of Bessel rings. Moreover, the method we proposed here can generate all orders and all types of skyrmions. Nonetheless, it may extend to construct skyrmion lattices [19,49] with self-healing and non-diffraction properties by the superposition of suitable Bessel modes. It is also worth noticing that our experimental configuration can generate all possible order optical bimerons in the absence of quarter wave plate present after second PBS [50]. The simplicity and compactness of the present experimental configuration can be easily interfaced with any device application. Also, our method of generation can be used to generate skyrmions in a wide range of electromagnetic spectra. The optical skyrmions generated in LG modes may have a change in their polarization structure while they are propagating. However, optical skyrmions produced in Bessel modes have a constant polarization structure upon propagation with non-diffraction and self-healing features. The Bessel skyrmions can have unique 3D polarization structures under high numerical aperture diffractive elements. Therefore, another interesting direction would be the investigation of 3D polarization textures in Bessel skyrmions for their applications in micro and nanoscale [51,52]. The skyrmions generated in the present context are 2D skyrmions and have the spin texture in the 2D plane (transverse plane). However, our analysis may open the door to investigate the 3D skyrmions called hopfions [53,54] which have 3D skyrmion texture in the volume formed by the transverse plane and longitudinal position. Optical skyrmions generated in the Bessel profiles may be useful in material processing to create 3D structures, optical data storage in terms of polarization, cryptography for secure communications [55-57].

## 6. Conclusion

We report optical skyrmions in the Bessel profile which can have non-diffraction and self-healing properties. The skyrmion textures created in Bessel profiles have completely different structural and propagation properties with reference to the skyrmions generated in LG modes. In order to utilize these optical skyrmions for applications in experimental laboratories, we proposed a simple and compact method for their experimental realization. We can easily construct these optical skyrmions with high output power in any time scale of laser pulses based on a couple of SPPs and one axicon with the aid of a Mach-Zander interferometer. We anticipate that such generated high power optical skyrmions in any pulsed laser source will have utility for applications in material processing, and optical engineering. We can easily transfer the state of polarization to generate all types of skyrmions by adjusting the predefined parameters of the experimental configuration. In the present context, the order of the skyrmions depends on the order of LG mode and their mode matching governed by the order of HoG mode. The skyrmions generated based on this technique will not have any on-axis intensity modulations due to the round/blunt tip of the axicon. The range and peak position of the skyrmions along the propagation axis depends on their order. Furthermore, we discussed possible artifacts originated in skyrmions while they are synthesized by the superposition of two higher order Bessel beams.